\documentclass[conference]{IEEEtran}
\usepackage{epsfig}
\usepackage{amsmath} 
\usepackage{gensymb}
\usepackage{siunitx}
\usepackage[center]{caption}
\def\BibTeX{{\rm B\kern-.05em{\sc i\kern-.025em b}\kern-.08em
    T\kern-.1667em\lower.7ex\hbox{E}\kern-.125emX}}

\usepackage{multicol}
\usepackage{color}
\usepackage{datetime}
\newcommand{\red}{\color {red}}

\newdateformat{mydate}{14$^{th}$ 
\monthname[4]  \the\year}

\title{\LARGE \bf
A simple Stochastic SIR model for COVID-19 Infection Dynamics for Karnataka after interventions -- Learning from  European Trends  \\\vspace*{20pt} \normalsize  
\mydate\today}

\author{\IEEEauthorblockN{Ashutosh Simha}
\IEEEauthorblockA{\textit{Department of Software Science} \\
Tallinn University of Technology, Estonia \\
ashutosh.iisc@gmail.com}
\and
\IEEEauthorblockN{R. Venkatesha Prasad}
\IEEEauthorblockA{\textit{Delft University of Technology}\\
Delft, the Netherlands \\
rvprasad@acm.org}
\and
\IEEEauthorblockN{Sujay Narayana}
\IEEEauthorblockA{\textit{Delft University of Technology}\\
Delft, the Netherlands \\
sujaynarayana@gmail.com}
}

\begin{document}

\maketitle

\thispagestyle{empty}
\pagestyle{empty}

\begin{abstract}
In this short note we model the region-wise trends of the evolution to COVID-19 infections using a stochastic SIR model. The SIR dynamics are expressed using \textit{It\^o-stochastic differential equations}. We first derive the parameters of the model from the available daily data from European regions based on a $'x'$-days\footnote{To find the value of $'x'$, please look at the graphs. This is because, this article is constantly being updated with new data every 2-3 days.} history of infections, recoveries and deaths. The derived parameters have been aggregated to project future trends for the Indian subcontinent, which is currently at an early stage in the infection cycle. The projections are meant to serve as a guideline for strategizing the socio-political counter measures to mitigate COVID-19. This article considers the latest data for Europe and India. 
\end{abstract}

\section{Introduction}
COVID-19 has brought misery to many regions in the world, especially in Europe. However, it has just taken off in other parts of the world. The remaining regions, including India hosts huge population approximately more than 50\% of the world population. There is an eminent danger and most of these population are under the poverty-line or the countries do not have huge resources for medical interventions for affected patients. The options are limited -- either go for herd immunity which can backfire and it is too risky or lockdown and restrict the population from moving around. Thus we try to factor in the percentage of population allowed to move around (some are essential services and some not obeying the orders) and assuming the mobility factors appropriately\footnote{For example, we assume that a person is mobile only around (60 x 20)\,km$^2$ area, which is approximately 10\% of the area of Karnataka. Such assumption which on the surface looks admissible}, we try to predict what can happen during the lockdown. We specifically take Italian case for Karnataka because of the population resemblance.

The dynamics governing the evolution of the COVID-19 infections have been modeled using a stochastic differential equation SIR model \cite{sir}. The parameters of this model have been initially optimized for a set of European regions, individually. An aggregate of these parameters have then been used for projecting the future trends for the Indian region, specifically for Karnataka state. Multiple projections have been generated by varying the exposure factor $E_f$ that influences the growth rate of infections. This is done in order to provide insights for selective quarantining and lockdowns.

\section{Stochastic SIR model}
The evolution of COVID-19 infections in each region, has been modeled via the stochastic susceptible-infected-recovered (SIR) model \cite{sir} which is given as,
\begin{eqnarray}
d{S}(t)&=&-E_f\beta S(t)C(t)dt \nonumber \\
d{C}(t)&=&\big(E_f\beta S(t) C(t)-\gamma C(t)\big)dt+ \sigma C(t)dW_t\nonumber  \\
dR(t)&=&\gamma C(t)dt-\sigma C(t)dW_t,
 \\ \nonumber  \\
C(0)&=&C_0,~ S(0)=S_0 \nonumber \\
S(0)&=&P_{total}-R(0)-C(0).
\end{eqnarray}

\subsection*{Model states}
\begin{enumerate}
    \item $t$ is a daily-time parameter.
    \item $C(t)$ denotes the number of active infections at time $t$. 
    \item $S(t)$ denotes the total susceptible population at time $t$.
    \item $R(t)$ denotes the total number of recoveries and deaths at time $t$.
    \item $dC(t), dS(t), dR(t)$ denotes the change in the states at time $t$.
    \item $dW_t$ is an incremental Weiner process (Brownian motion), which models the randomness in the evolution. 
    \item $E_f\in[0,1]$ which multiplies $\beta$ is the exposure factor  which models the interventions to subdue infection spread, such as lockdowns, quarantining, and preventive measures. 
    \end{enumerate}

\subsection*{Model parameters}
\begin{itemize}
    \item Growth rate: the constant $\beta$ denotes the growth rate, which factors the rise in the number of infections, due to interactions between susceptible and infected population. This parameter is a lumped constants which is meant to account for: (a) the population size, (b) reproduction number $R_0$ of COVID-19, and (c) exposure-factor (which depends on mobility, precautionary measures, etc.). 
    \item $\gamma$ is the rate of outcomes, i.e., the rate at which the infections are neutralized, which may be due to recovery or death. It is assumed that recovered persons would not spread the infections again (at least for a window of a month). 
    \item $\sigma$ is a parameter used to model the stochasticity or randomness in the evolution, which may cause local deviations from the typical (exponential) trends.
   \item $P_{total}$ is the population of the region, $C_0$ and $S_0$ are initial number of infections and susceptible individuals.
\end{itemize}
\vspace{0.5cm}

\section{Parameters based on European trends}
The parameters of the SIR model were optimized based on the data obtained for different European regions and India. The criterion for optimization was to simultaneously minimize the square integral error, terminal error and terminal rate error, between the actual data and  daily samples of the simulated data. Further, because we have data for more number of days for European countries, we try to use the parameters from those countries and appropriate it on the Indian data which is for lesser number of days. The assumption is that India may be in the catch game (which, of course, we do not want) if the behaviour of people is taken  to be similar. We have to base our predictions based on some gross assumptions under the given circumstances. 

\textit{Note:}
 We have taken the Italian and German data $4$ days earlier than France and Spain in order to reflect the earlier trend before their respective lockdown conditions. This will enable us to simulate for various levels of lockdown percentages (exposure factor).

\begin{center}
     \begin{table}
     \caption{Model parameters assumed. These numbers are slightly pessimistic.}

    \begin{tabular}{|c|c|c|c|c|}
     \hline
    Region & $P_{total}$ & $\beta$ & $\gamma$ & $\sigma$\\
    \hline 
    Germany & $8.28\times 10^7$ & $4.6\times 10^{-9}$ & $0.005(1-28),$ & $0.01$  \\
    ~&~&~&$0.045(29-39)$&~\\
    Spain & $4.67\times 10^7$ & $7.5\times 10^{-9}$ & $0.06$ & $0.02$  \\

    France & $6.7\times 10^7$ & $4.4\times 10^{-9}$ & $0.04$ & $0.02$  \\ 
    Italy & $ 6.055\times10^7$ & $4.5\times10^{-9}$ & $0.04$ & $0.025$  \\ 
    India & $133.92\times 10^7$ & $1.75\times 10^{-10}$ & $0.02$ & $0.01$  \\
    \hline
   
    \end{tabular}
    \end{table}
\end{center}
\vspace{0.5cm}

\section{Simulation results}
The SIR model has been simulated and the parameters have been optimized based on the infection trends obtained for European countries and India (after 15$^{th}$ March, when the infections started to show an exponential trend). The stochastic differential equations have been simulated using the Euler-Maruyama numerical integration method \cite{sde}. We present below how our model follows the statistics from  various European countries.
 \begin{figure}[h]
 	\centering
     	\includegraphics[width=0.95\linewidth,height=0.76\linewidth]{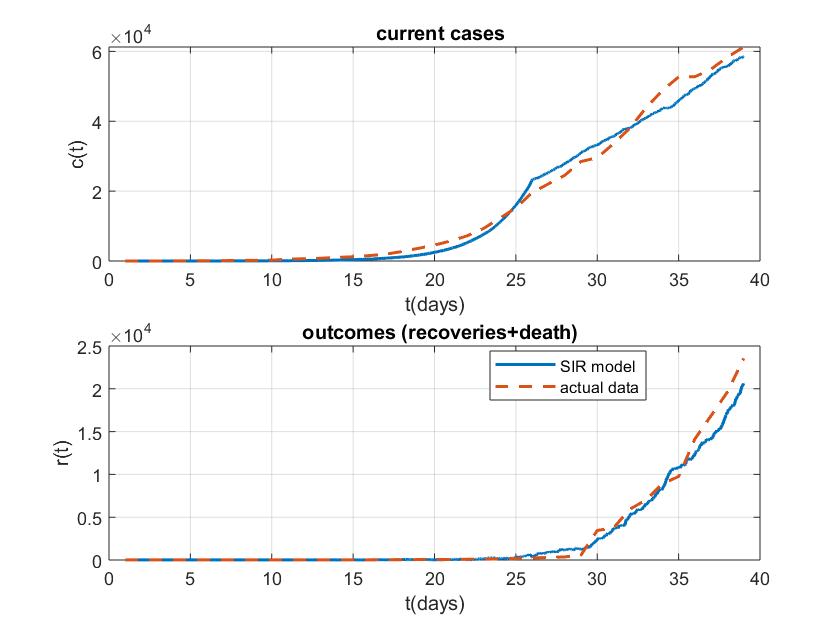}
    	\caption{SIR model for Germany. Day $1=24^{th}$ Feb 2020. Interventions after day 26 resulted in $E_f=0.28$}
    	\label{Fig: germany}
    	\end{figure}
 \begin{figure}[h]
    \centering
    	\includegraphics[width=0.95\linewidth,height=0.76\linewidth]{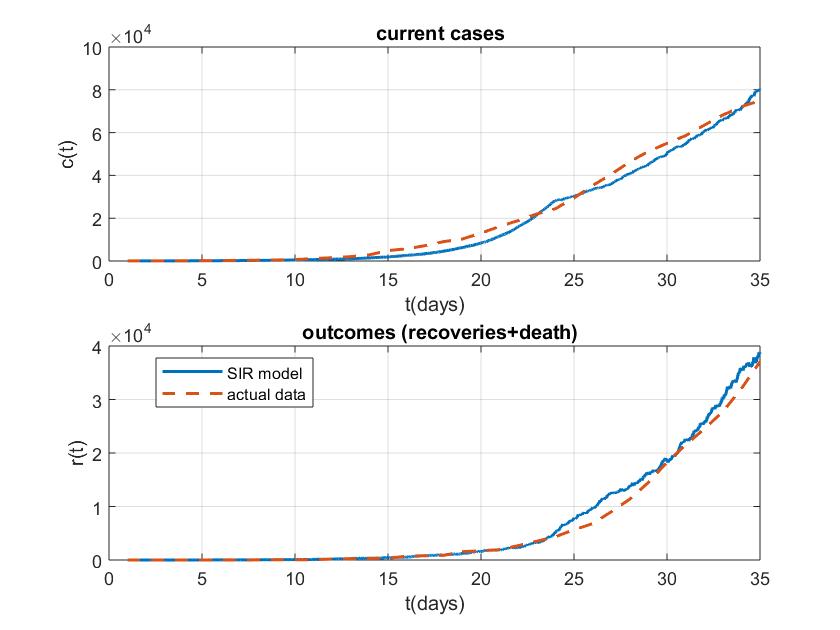}
    	\caption{SIR model for Spain. Day $1=28^{th}$ Feb 2020. Interventions after day 23 resulted in $E_f=0.45$}
    	\label{Fig: spain}
    \end{figure}
    
 \begin{figure}[h]
        \centering
    	\includegraphics[width=0.95\linewidth,height=0.76\linewidth]{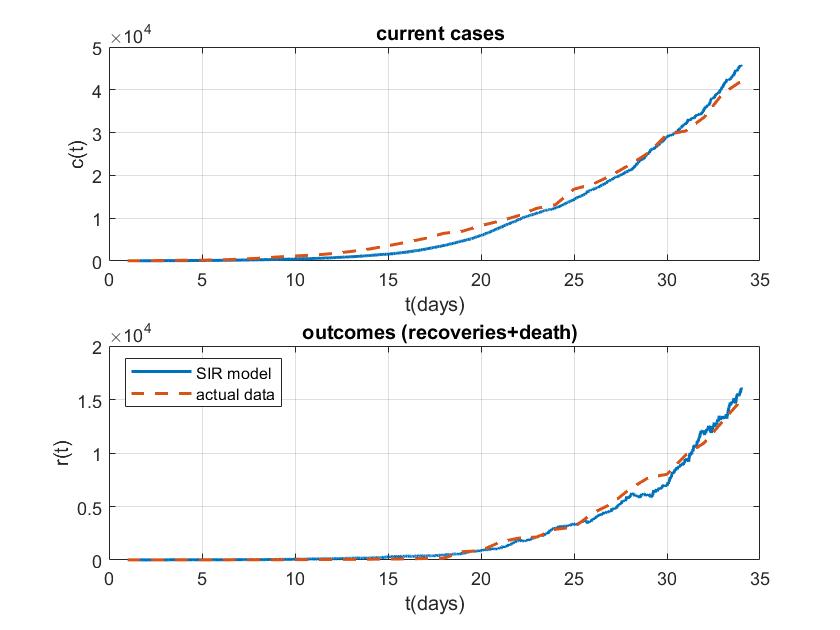}
    	\caption{SIR model for France. Day $1=28^{th}$ Feb 2020. Interventions after day 21 resulted in $E_f=0.6$}
    	\label{Fig: France}
 \end{figure}   	
\begin{figure}[h]
        \centering
    	\includegraphics[width=0.95\linewidth,height=0.76\linewidth]{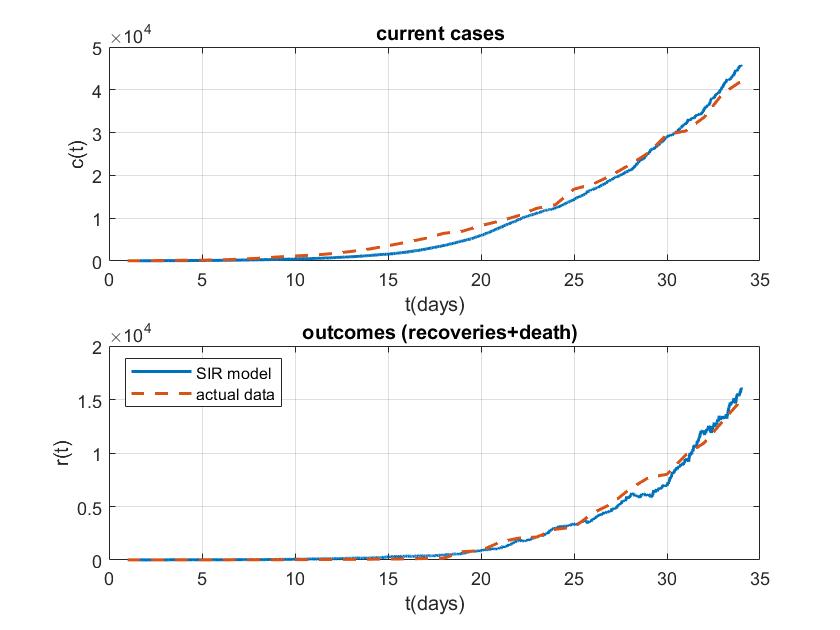}
    	\caption{SIR model for Italy. Day $1=24^{th}$ Feb 2020. Interventions after day 19 resulted in $E_f=0.45$}
    	\label{Fig: italy}
 \end{figure} 
 
  \begin{figure}[t]
     	\centering
    	\includegraphics[width=0.95\linewidth,height=0.76\linewidth]{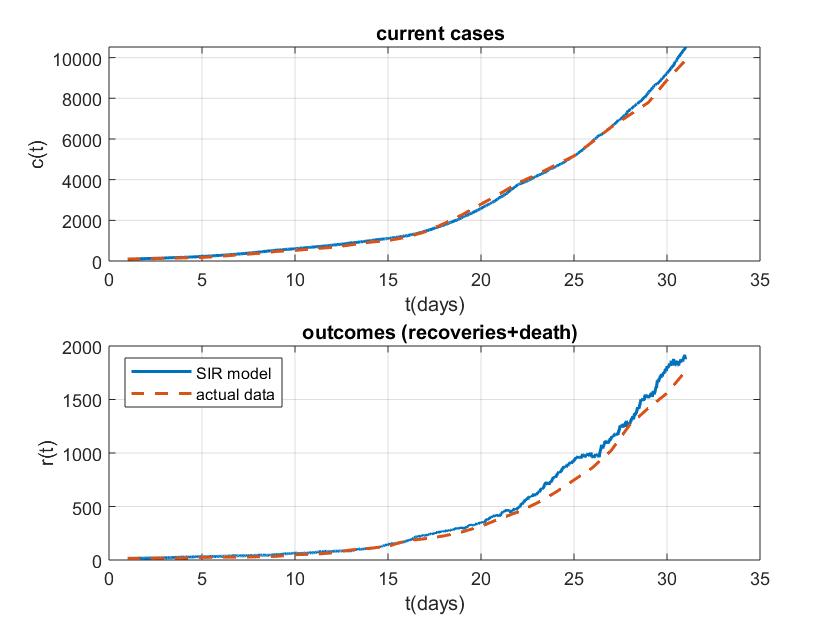}
    	\caption{SIR model for India. Day $1=15^{th}$ March 2020.  Intervention: day 8 to day 15; $E_f=0.6$, day 15 to day 22; $E_f=0.6$ (lockdown violations), day 22 onwards $E_f=0.6$. Projections are done from day 23 onwards (when the parameters were fixed) in order to assess the validity of the model. This figure will be continuously updated.}
    	\label{Fig: india}
    \end{figure}
\vspace{0.5cm}


 \section{Projections for India}
 \subsection{Nationwide projection}
  The SIR model has been simulated beyond the $`x'$-days of considered data for India, with parameters averaged over the European regions considered above, as well as with the parameters obtained for the Indian region. This is done in order to demonstrate the possible growth of infections in India, based on the current trend so far (which is an optimistic case), or based on the European trend (which is the case to be prepared for). The exposure factor $E_f$ has been varied to scale down the rate of infections $\beta$.  This factor illustrates the additional reduction in mobility due to state-wide lock-down and quarantining measures, which are essential for maintaining new infections within reasonable limits. For India we have two sets, (i) using the available data only from India and (ii) taking the model parameters from considered European countries and projecting on Indian case.
  
  \textit{Note:} We have four graphs in each set. For example, when we use $E_f=1$, that means there is no lockdown. When we use $E_f=0.5$ exposure half of the country/state is in lockdown. This is to show the gravity of the situation if the lockdown is not respected.
  
 \begin{figure}[h]
 	    \centering
    	\includegraphics[width=0.95\linewidth,height=0.76\linewidth]{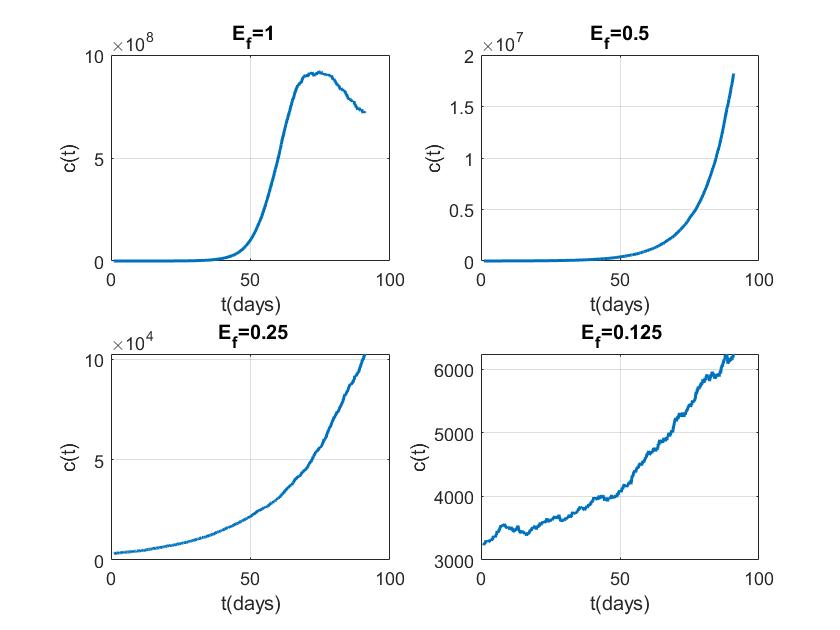}
    	\caption{Projections for India with varying rates of exposure using Indian parameters. Since we already have the data for previous days, our prediction was from the Day-$1=3^{rd}$ April.}
    	\label{Fig: india_indiaparams}
    \end{figure}
    
     \begin{figure}[h]
     	\centering
    	\includegraphics[width=0.95\linewidth,height=0.76\linewidth]{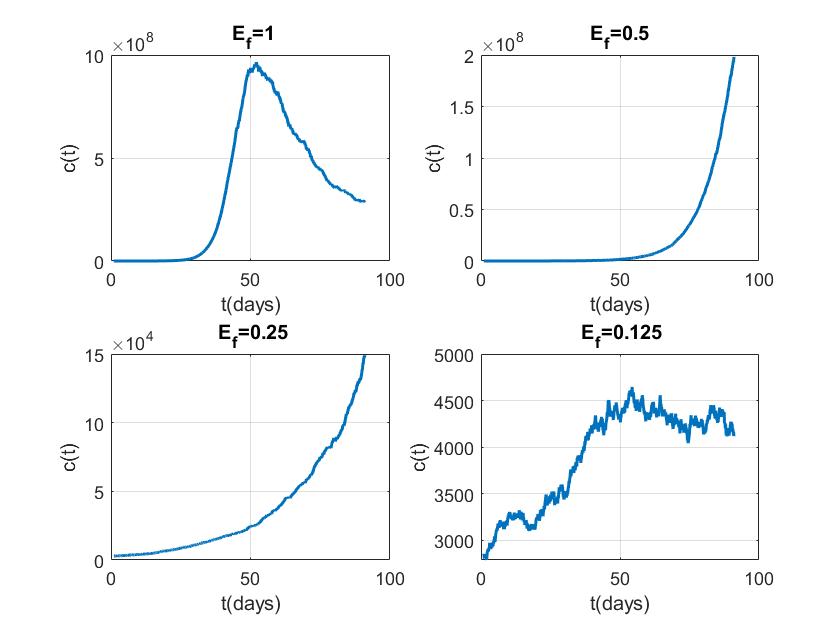}
    	\caption{Projections for India with varying rates of exposure using Average European parameters. Since we already have the data for previous days, our prediction was from the Day-$1=3^{rd}$ April.}
    	\label{Fig: india_avgparams}
    \end{figure}
   
    
    \subsection{Karnataka Projections}
The  average European parameters have been used to project the future trends for the state of Karnataka.  The initial number of infections is assumed taken as 113 (3rd April) to account for additional unreported or undetected infections, over reported ones. We have made the assumption here that lockdown is in place and we have taken varied percentage of people obeying the rules. 

\begin{figure}[t]
    	\centering
    	\includegraphics[width=0.95\linewidth,height=0.76\linewidth]{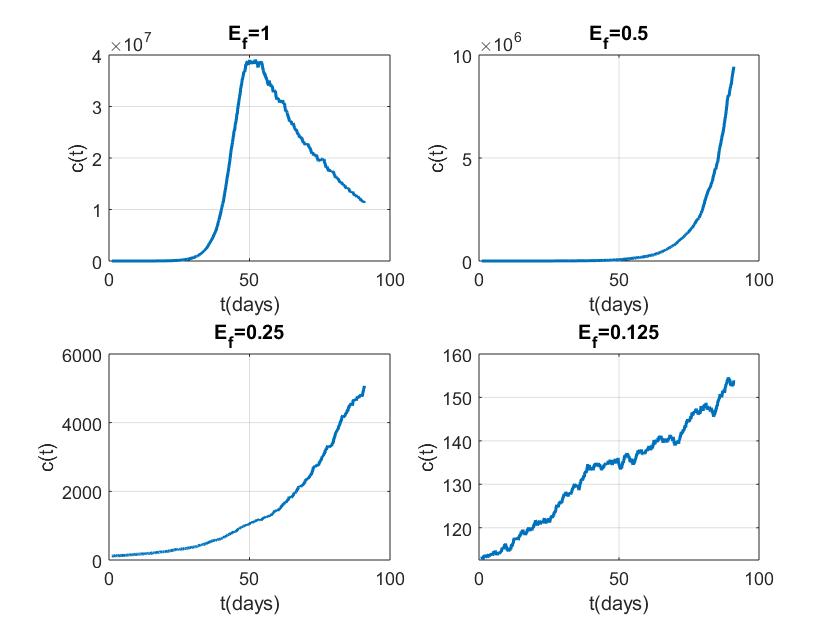}
    	\caption{Projections for Karnataka with varying rates of exposure using averaged European parameters. Initial  active infections taken as 113 as on $3^{rd}$ April. The random fluctuations are dominant in the fourth quadrant, due to lower $E_f$.}
    	\label{Fig: karnataka_avgparams}
    \end{figure}
    
    \begin{figure}[tbh]
    	\centering
    	\includegraphics[width=0.95\linewidth,height=0.76\linewidth]{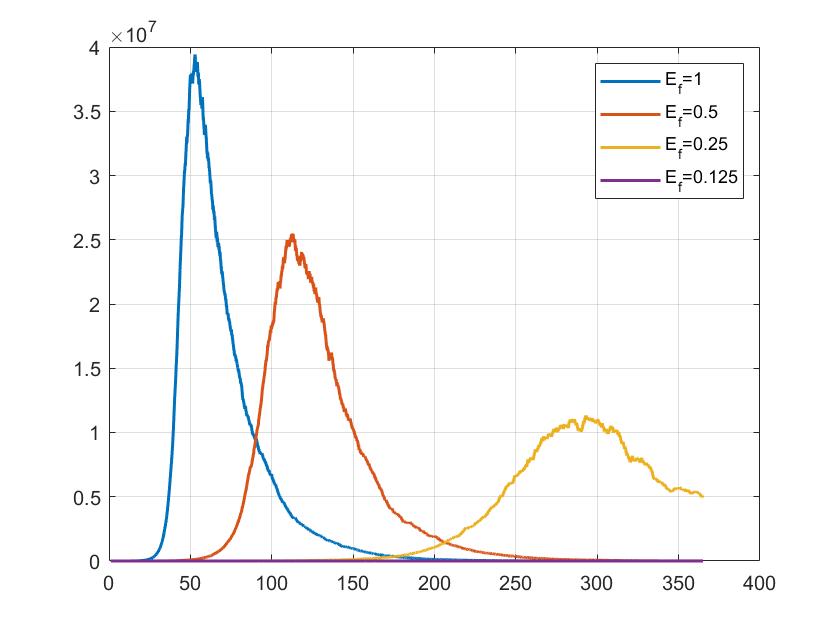}
    	\caption{Projections for Karnataka for $365$ days with varying interventions using European average parameters.}
    	\label{Fig: karnataka365}
    \end{figure}
Let us summarize very briefly what we observe.
\begin{itemize}
           \item Below $25\%$ exposure required for avoiding exponential growth for Karnataka, as well as for India using both Indian parameters and averaged European parameters.
               \item  Fig.\ref{Fig: karnataka365}  suggests the inevitable need for an \textbf{Extended lockdow period for several months} for Karnataka, or until a vaccine is available. 
        \item By `lockdown' we mean restricted mixing i.e. fewer susceptible individuals $s(t)$ interact with fewer infected individuals $c(t)$, thereby introducing a reduction in the factor $E_f$ in the dynamics. 
        {\red \item \textbf{Note:} The previous version of this draft uploaded on arxiv on 29th march (v2) predicted a total number of $10080$ active infections in India, 30 days from 15th of March (See Fig.8 of v2 of \cite{simha_vp}). As on 14th April, the number of \textit{recorded} active infections is $9872$. The number of active infections on 29th March was $902$.}
\end{itemize}

\section{Varying Lockdown Schedules}
\begin{figure}[tbh]
    	\centering
    	\includegraphics[width=0.95\linewidth,height=0.76\linewidth]{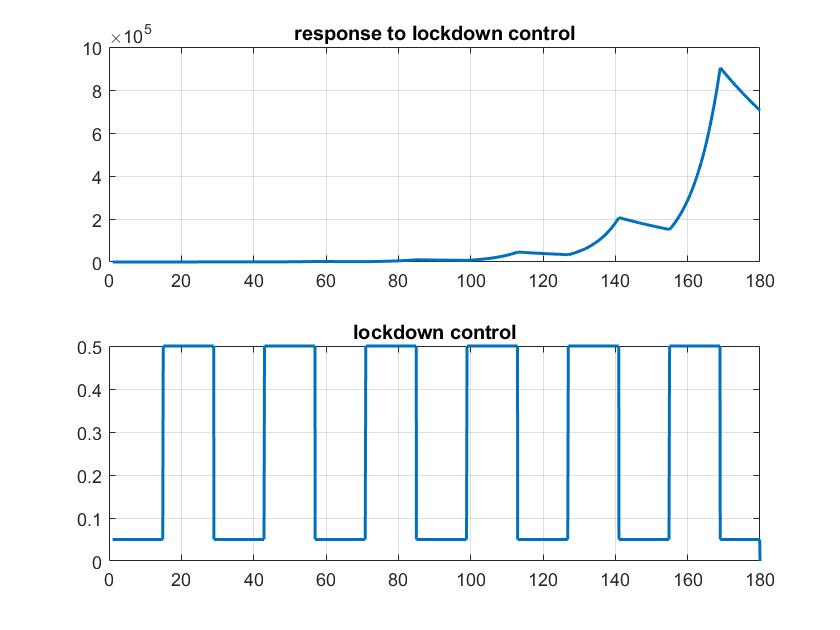}
    	\caption{For Karnataka with initial condition of 113 infections on 3rd of April 2020. 95\% Lockdown for two weeks and 50\% relaxation (for example schools are closed throughout, essential businesses are open).}
    	\label{Fig: lock2}
    \end{figure}

\begin{figure}[tbh]
    	\centering
    	\includegraphics[width=0.95\linewidth,height=0.76\linewidth]{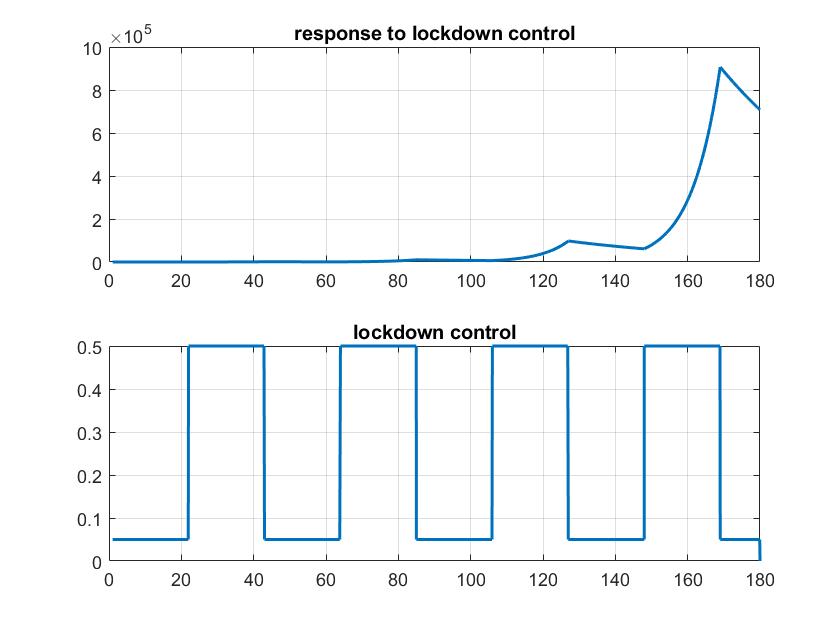}
    	\caption{3 weeks cycle similar to Fig.\ref{Fig: lock2}.}
    	\label{Fig: lock3}
    \end{figure}

\begin{figure}[tbh]
    	\centering
    	\includegraphics[width=0.95\linewidth,height=0.76\linewidth]{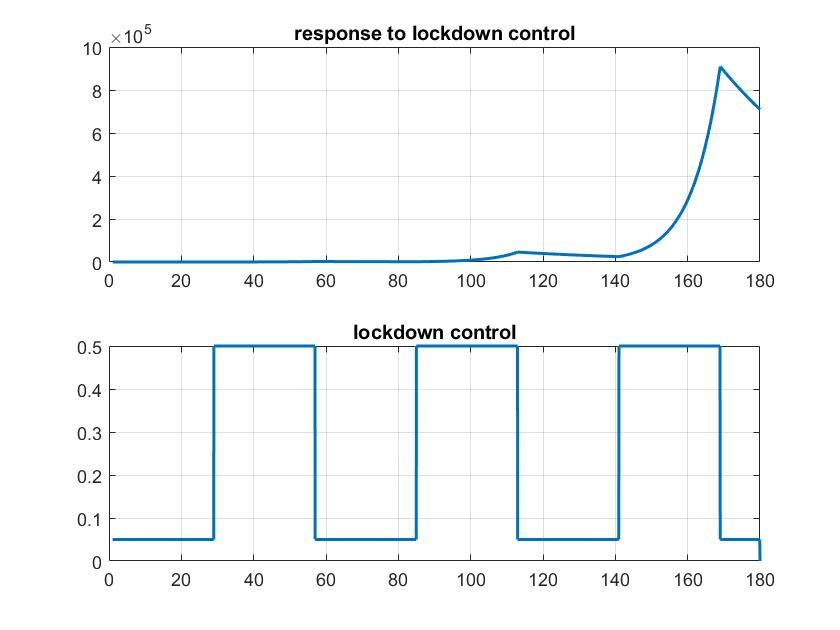}
    	\caption{4 weeks cycle similar to Fig.\ref{Fig: lock2}.}
    	\label{Fig: lock4}
    \end{figure}

\begin{figure}[tbh]
    	\centering
    	\includegraphics[width=0.95\linewidth,height=0.76\linewidth]{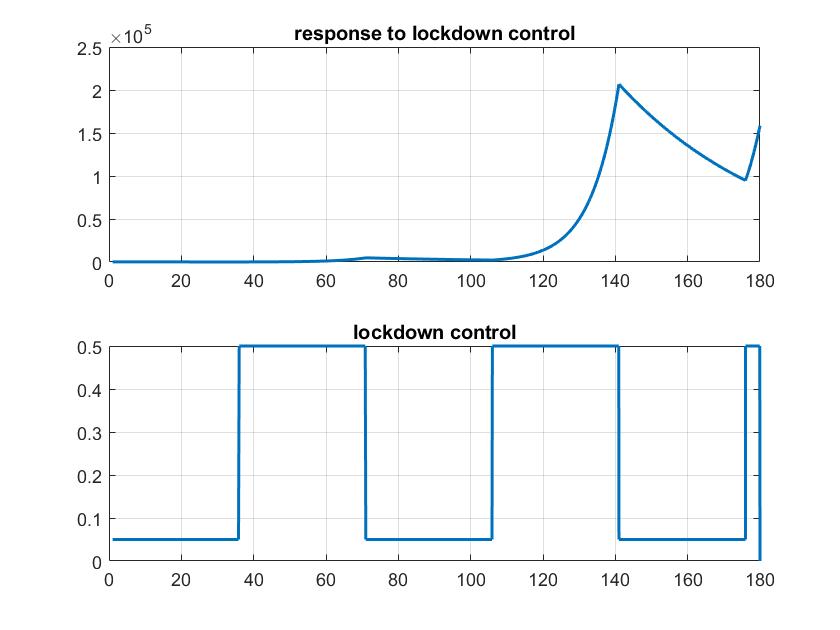}
    	\caption{5 weeks cycle similar to Fig.\ref{Fig: lock2}.}
    	\label{Fig: lock5}
    \end{figure}

\begin{figure}[tbh]
    	\centering
    	\includegraphics[width=0.95\linewidth,height=0.76\linewidth]{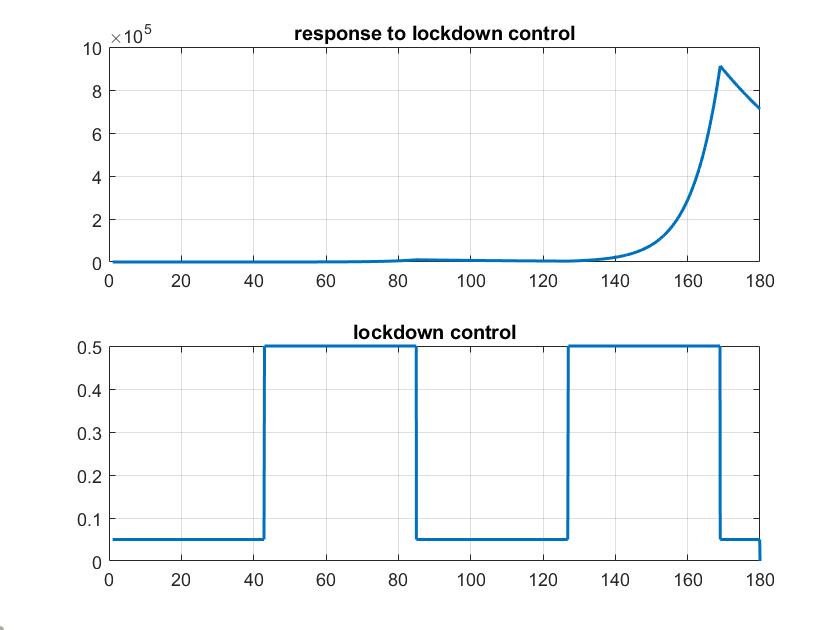}
    	\caption{6 weeks cycle similar to Fig.\ref{Fig: lock2}.}
    	\label{Fig: lock6}
    \end{figure}

\begin{figure}[tbh]
    	\centering
    	\includegraphics[width=0.95\linewidth,height=0.76\linewidth]{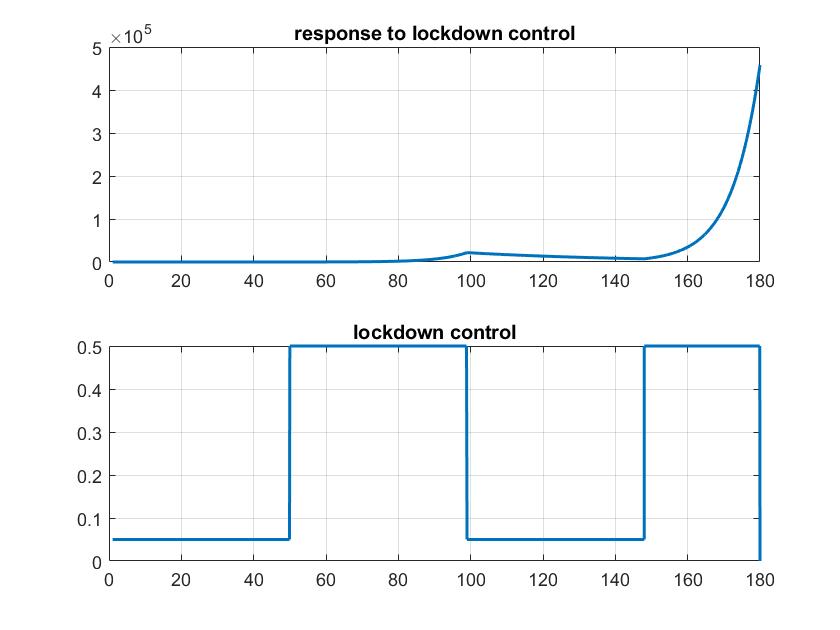}
    	\caption{7 weeks cycle similar to Fig.\ref{Fig: lock2}.}
    	\label{Fig: lock7}
    \end{figure}

\begin{figure}[tbh]
    	\centering
    	\includegraphics[width=0.95\linewidth,height=0.76\linewidth]{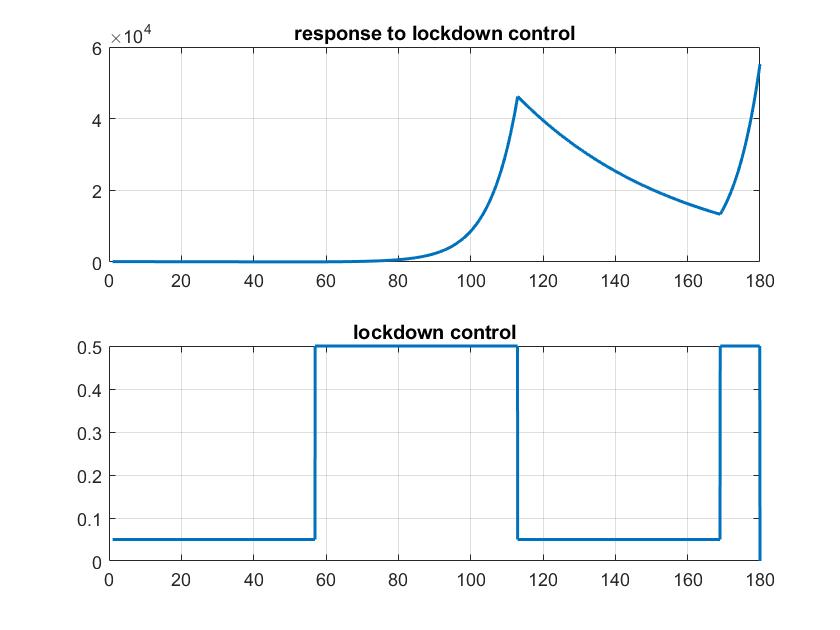}
    	\caption{8 weeks cycle similar to Fig.\ref{Fig: lock2}.}
    	\label{Fig: lock8}
    \end{figure}

The usual tendency is to relax a bit when the situation is under control to address the economic issues. {\red{However in the case of COVID-19, we can not do that because we would be just postponing the issues. This we can see in the Figs \ref{Fig: lock2} to \ref{Fig: lock8}.}} To understand what can happen if the lockdown is relaxed by 50\% for sometime and again the lockdown is imposed. For example, we close all the schools, restaurants and theaters and allow only workers to move out. This idea looks like a good compromise since essential work and people having difficulties because of lockdown can carryout their work. Further, only 50\% of the population are allowed. To understand such a situation, we simulated the ON-OFF model with 50\% duty cycle. 

Simulations are done by varying $E_f$ over time, in order to understand the infection response to time varying lockdown schedules over 6 months for Karnataka. Note that a lower $E_f$ could also indicate a mobile society, but with appropriate preventive measures, not just purely lockdown. 

The various response curves have been plotted to serve as a guideline for selecting a particular lockdown schedule, or a mix of schedules, such that the infections are minimized simultaneously with the socio-economic losses incurred due to the lockdown. However, consistently the data shows that the effect of such a relaxation will  only to shift the peaking and it does not help in any ways except getting a few  months for preparation initially. We know that considering the scales for India we will not be able to handle even after six months the numbers that we see from the simulations. The trend is that after low infections over initial cycles, the infections will fluctuate with large magnitude. This method could be adopted if good medical support can be guaranteed. A similar simulation can be found in \cite{mit}. Our results Figs \ref{Fig: lock2} to \ref{Fig: lock8} are for Karnataka with initial condition of 113 infections on 3rd of April 2020.

Thus the only concrete step is to lockdown till we localize the cases with 99.999\% accuracy and then allow people to cautiously start resuming their work. 

    \begin{figure}[tbh]
    	\centering
    	\includegraphics[width=0.95\linewidth,height=0.76\linewidth]{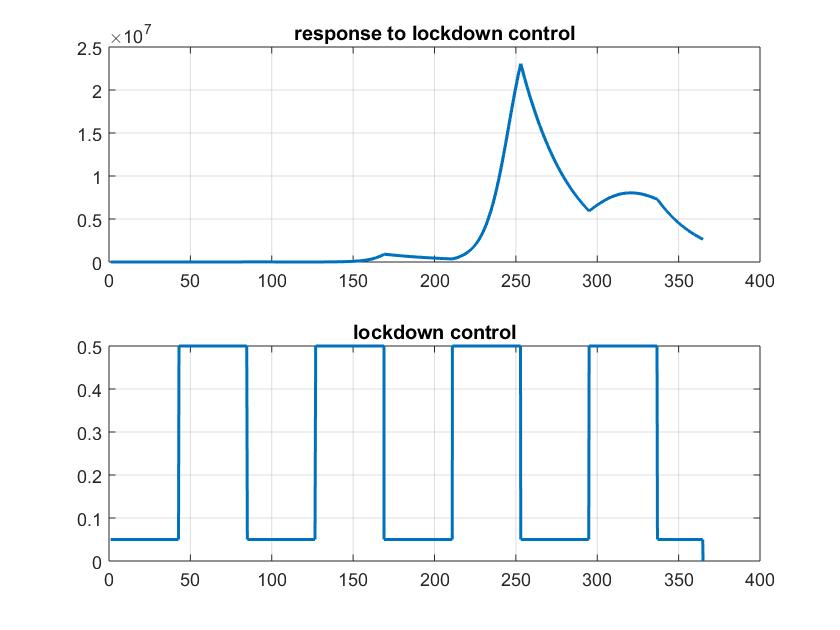}
    	\caption{6 weeks cycle for one year.}
    	\label{Fig:lock365_6}
    \end{figure}
      \begin{figure}[tbh]
    	\centering
    	\includegraphics[width=0.95\linewidth,height=0.76\linewidth]{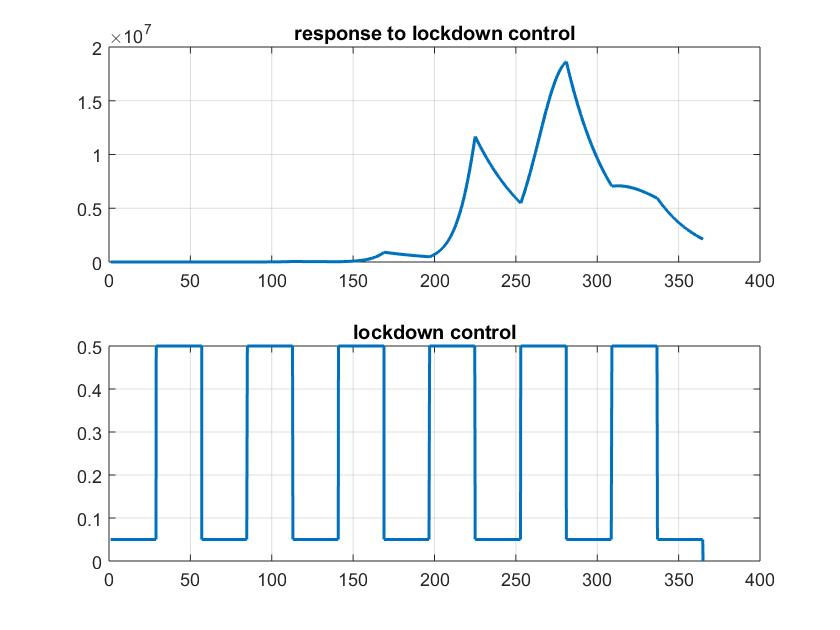}
    	\caption{4 weeks cycle for one year.}
    	\label{Fig:lock365_4}
    \end{figure}

\section{Conclusions}
This is a continuous work in which we are trying to find the  model parameters everyday and project the possible scenarios, by varying the exposure factor for the rate of infection, as a result of evolving levels of quarantining. While this is not a completely verifiable projection, the model parameters look quite consistent, however they may reflect an over estimation of the projected number of infections in order to compensate for the unreported or undetected infections. Thus, we may take the numbers in this note to be a guide for further action by the law enforcing authorities. 


\begin{thebibliography}{}
\bibitem{sir}
Maki, Yoshihiro, and Hideo Hirose. \textit{"Infectious disease spread analysis using stochastic differential equations for SIR model."} 2013 4th International Conference on Intelligent Systems, Modelling and Simulation. IEEE, 2013.

\bibitem{sde}
Kloeden, P. E., and Pearson, R. A. (1977). \textit{The numerical solution of stochastic differential equations.} The ANZIAM Journal, 20(1), 8-12.

\bibitem{simha_vp}
Simha, Ashutosh, R. Venkatesha Prasad, and Sujay Narayana. "A simple Stochastic SIR model for COVID 19 Infection Dynamics for Karnataka: Learning from Europe." arXiv preprint arXiv:2003.11920 (2020), v2(29th march).
\bibitem{mit}
[Online] Gideon Lichfield. "We're not going back to normal." https://www.technologyreview.com/2020/03/17/905264/coronavirus-pandemic-social-distancing-18-months/  [accessed on 15th April 2020]
\end{thebibliography}
\end{document}